\documentclass[useAMS,usenatbib]{mn2e}
\usepackage{graphicx}
\usepackage{epsfig}

\begin{document}
\title[IGR J16465--4507]{The orbital period in the supergiant fast X-ray transient IGR J16465--4507}
\author[D. J. Clark et al.] {D. J. Clark$^{1,2}$\thanks{E-mail: djc400@soton.ac.uk}, V. Sguera$^{3,4}$, A.J Bird$^{1}$, V. A. McBride$^{1}$, A. B. Hill$^{5}$, S. Scaringi$^{6}$, \newauthor
S. Drave$^{1}$, A. Bazzano$^{4}$, A.J Dean$^{1}$\\
$^{1}$School of Physics and Astronomy, University of Southampton, Southampton. SO17 1BJ, UK\\
$^{2}$Centre dEtude Spatiale des Rayonnements, CNRS/UPS, BP 4346, 31028 Toulouse, France\\
$^{3}$INAF, Istituto di Astrofisica Spaziale e Fisica Cosmica, Via Gobetti 101, I-40129 Bologna, Italy\\
$^{4}$INAF, Istituto di Astrofisica Spaziale e Fisica Cosmica, Via Fosso del Cavaliere 100, I-00133 Rome, Italy\\
$^{5}$Laboratoire d'Astrophysique de Grenoble, UMR 5571 CNRS, Universit\'e Joseph Fourier, BP 53, 38041 Grenoble, France\\
$^{6}$Department of Astrophysics, IMAPP, Radboud University Nijmegen, P.O. Box 9010, 6500 GL Nijmegen, The Netherlands
}
\date{Accepted 2010 May 24. Received 2010 May 24; in original form 2010 March 2}
\pagerange{\pageref{firstpage}--\pageref{lastpage}} \pubyear{2010}
\maketitle
\label{firstpage}
\begin{abstract}
Timing analysis of the {\it INTEGRAL}-IBIS and \emph{Swift}-BAT light curves of the Supergiant Fast X-ray Transient (SFXT) IGR J16465--4507 has identified a period of 30.32$\pm$0.02\,days which we interpret as the orbital period of the binary system.  In addition 11 outbursts (9 of which are previously unpublished) have been found between MJD 52652 to MJD 54764, all of which occur close to the region of the orbit we regard as periastron. From the reported flux outbursts, we found a dynamical range in the interval $\sim$30-80. Although in this regard 
IGR J16465--4507 cannot be considered a  classical SFXT for which typical dinamical ranges are $>$100, still our reported values are significantly
greater than  that of classical persistent variable supergiant HMXBs ($<20$), supporting  the idea that 
IGRJ16465--4507 is an intermediate SFXT system, much like few other similar cases 
reported in the literature.

\end{abstract}

\section{Introduction}
During the last few years, the IBIS instrument \citep{Ubertini:2003kn} on board the \emph{INTEGRAL} gamma-ray satellite \citep{Winkler:2003hh} has played a key role in discovering a previously unrecognized class of high mass X-ray binaries which display a very unusual fast X-ray transient behaviour and host a massive OB supergiant star as companion: the so-called Supergiant Fast X-ray Transients (SFXTs; \citealt{Zand:2005dp,Sguera:2005os,Sguera:2006ns,Negueruela:2006lt}). To date approximately 10 show clear characteristics of being a SFXT (plus the same number of candidates), characterized by long periods of low X-ray activity level (with luminosities or upper limits typically in the range 10$^{32}$-10$^{34}$ erg s$^{-1}$), interspersed with short, strong, flares lasting from a few hours to no more than a few days and reaching peak luminosities of 10$^{36}$-10$^{37}$ erg s$^{-1}$. The identification of periodicity in the SFXT outburst behaviour, most likely representing the orbital period, is a key diagnostic for studying the geometry of these systems and hence for understanding the still unknown physical mechanisms behind their very unusual behaviour (see \citealt{2009AdSpR..43.1464S} for a review). Because the fast outbursts from SFXTs are relatively rare occurrences, the discovery of such periodicities is a very challenging task which has been fulfilled only for an handful of SFXTs, mainly  through long-term temporal monitoring by \emph{INTEGRAL} (e.g. \citealt{2009MNRAS.399L.113C,Bird:2009yq, 2009MNRAS.397L..11J, 2009A&A...493L...1Z, 2007A&A...467..249S, 2007A&A...476.1307S}). 

IGR J16465--4507 classification as an SFXT is debatable. Its initial discovery on 6 September 2004 was triggered by the \emph{INTEGRAL} observation in the Norma region of a faint $\sim$9 mCrab source (18-60 keV) which produced a strong $\sim$28 mCrab flare on 7 September 2004 before returning to its undetectable pre-flare level \citep{2004ATel..329....1L}, demonstrating its fast X-ray transient nature. However since then, no more outburst activity has been reported from the source and its temporal and spectral behaviour above 20\,keV is largely unknown. A \emph{XMM-Newton} observation performed a few days after the source discovery allowed the detection of high intrinsic absorption ($\sim$6$\times$10$^{23}$ cm$^{-2}$) as well as a $\sim$228\,s period which was interpreted as the spin period of  a neutron star \citep{Walter:2006fv,2005A&A...444..821L}. A single bright infrared object falls in the refined \emph{XMM-Newton} position \citep{2004ATel..336....1H}, optical and infrared spectroscopy classified it as a supergiant star located at a distance of $9.5^{+14.1}_{-5.7}$\,kpc, although different spectral types have been proposed (O9.5Ia by \citealt{2008A&A...486..911N} or B0.5I by  \citealt{Negueruela:2006lt, Rahoui:2008ig}). \citet{Walter:2007ww} reported a 22--50 keV quiescent flux of $7.7\times10^{-12}$\,erg\,cm$^{-2}$\,s$^{-1}$ and suggest that IGR J16465--4507 is not a classical SFXT (for which the dynamical range is $>$100, \citealt{2010arXiv1005.1995C}) but a kind of intermediate SFXT still having a dynamic range greater than that of a classical persistent variable supergiant HMXB (for which the dynamic range is $<$20) such as Vela X-1 \citep{2008A&A...492..511K}. In comparison a very recent paper by \citet{2010MNRAS.tmpL..63L} find no evidence of flaring activity from 
IGR J16465--4507 and inferred  a dynamical range $<$10.

Here we report a comprehensive temporal study of IGR J16465--4507, based on long term \emph{INTEGRAL} monitoring, which led to the discovery 
of 9 previously unpublished outbursts as well as the $\sim$30 days periodicity found by \citet{2010MNRAS.tmpL..63L}, which we interpret as the binary period.

\section{Data Analysis}

\begin{figure}
	\includegraphics[width=0.95\linewidth,height=0.7\linewidth,trim = 0mm 0mm 0mm 5mm, clip]{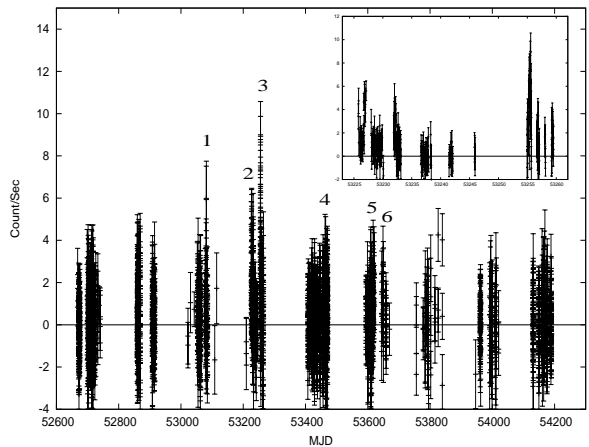}
	\caption{IBIS/ISGRI long term light curve (18-60\,keV) of IGR J16465--4507. Each data point represents the average flux during one ScW ($\sim$2000\,s). The six activity periods listed in table~\ref{table:oldbursts} are indicated in the light curve by the numbers from 1 to 6.  A zoomed view of the activity periods 2 and 3 (from MJD 53225.73 to 53259.48) is shown in the inset.}
	\label{figure:fulllc}

\end{figure}

Archival IBIS data from MJD 52652 to MJD 54764, giving an on source exposure of $\sim7.4$\,Ms, were processed with the {\it INTEGRAL} Off-line Science Analysis (OSA \citealt{2003A&A...411L.223G}) version 7.0.  A light curve for IGR J16465--4507 was generated on science window timescales ($\sim2000$\,s) over the 18-60\,keV energy range (fig.~\ref{figure:fulllc}).  The light curve was then searched iteratively for outbursts.  The shortest timescale region of the light curve that produces the highest detection significance greater than 4\,$\sigma$ is found.  This is the brightest outburst in the light curve.  This section of the light curve is then removed from the total light curve and the next highest significant section is searched for. 

\begin{table*}
	\centering
	\caption{Outbursts found in the 18--60\,keV ISGRI light curve, ordered by MJD. Peak luminosities have been calculated by assuming a distance of 9.5\,kpc. It is clear due to the separation between the outbursts found that several could be combined into periods of flaring activity, indicated in column 1.}
	\begin{tabular}{lr@{.}lcccr@{.}lr@{.}lcc}
	\hline
	Activity &  \multicolumn{2}{c}{Sig} 		& Start & Stop & Length &	 \multicolumn{2}{c}{Peak Flux} 	                                  & \multicolumn{2}{c}{Peak Luminosity} 	        & Orbital  &\\
	 Period	 &  \multicolumn{2}{c}{$\sigma$}	& MJD & MJD & Days    & \multicolumn{2}{c}{$10^{-11}$\,erg\,cm$^{-2}\,s^{-1}$} & \multicolumn{2}{c}{$10^{36}$\,erg\,s$^{-1}$} 	& Phase  &\\
	\hline
    1 & 	  4&05 & 	 53081.12 & 	53081.54 & 	0.40 & ~~~~~~~~~42&12	&~~~~~~~~4&52 & 	  0.24 & 	\\
    2 & 	  5&27 & 	 53225.73 & 	53226.56 & 	0.83 & 	  		 33&27 	& 		3&57 & 	  0.02 & 	\\
    2 & 	 14&03 & 	 53226.59 & 	53227.16 & 	0.54 & 	  		 39&31 	& 		4&22 & 	  0.04 & 	\\
    2 & 	  4&69 & 	 53227.88 & 	53229.90 & 	1.99 & 	  		 20&49 	& 		2&20 & 	  0.11 & 	\\
    2 & 	  7&71 & 	 53231.89 & 	53232.27 & 	0.36 & 	  		 35&10	& 		3&77 & 	  0.21 & 	\\
    3 & 	 16&97 & 	 53255.24 & 	53255.67 & 	0.42 & 	  		 63&18	& 		6&79 & 	  0.98 & 	\\
    3 & 	  5&37 & 	 53259.19 & 	53259.48 & 	0.28 & 	  		 24&36	& 		2&62 & 	  0.11 & 	\\
    4 & 	  4&01 & 	 53468.18 & 	53471.82 & 	3.63 & 	  		 25&41	& 		2&73 & 	  0.06 & 	\\
    5 & 	  5&35 & 	 53617.27 & 	53618.64 & 	1.34 & 	  		 26&18	& 		2&81 & 	  0.93 & 	\\
    5 & 	  4&47 & 	 53620.30 & 	53620.91 & 	0.59 & 	  		 23&93 	& 		2&57 & 	  0.02 & 	\\
    6 & 	  4&43 & 	 53648.85 & 	53650.54 & 	1.68 & 	  		 25&27 	& 		2&71 & 	  0.98 & 	\\
	\hline
	\end{tabular}
	\label{table:oldbursts}
\end{table*}

\begin{figure}
	\includegraphics[width=0.95\linewidth,trim = 8mm 3mm 0mm 6mm, clip]{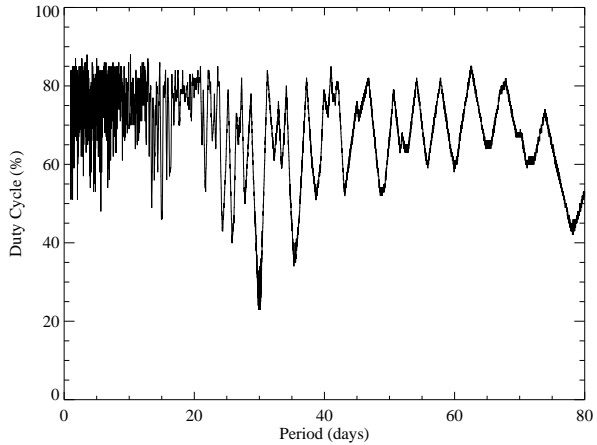}
	\caption{Phase Dispersion Minimization of the outbursts found shows a minimum at 29.88 days with a 23\% duty cycle.}
	\label{figure:PDM}
	\vspace{-6pt}
\end{figure}

In total 11 outbursts have been found in the light curve (Table~\ref{table:oldbursts}).  The outbursts range from short flares of 6.8\,h to periods of flaring activity up to 87 hours.  However, many of these outbursts are close together and if considered to be part of an overall period of flaring activity, only 6 periods of flaring remain.  By folding the outburst times on different periods, an estimation of the duty cycle required over the orbit to explain all the outbursts can be made.  This Phase Dispersion Minimization (PDM) technique produces a minimum of $\sim$23\% duty cycle at  a period of 29.88 days in our data (Fig.~\ref{figure:PDM}), which could indicate the orbital period and that outbursts occur over $\sim$23\% of the orbital phase. A Monte-Carlo simulation of the PDM performed by randomising the outburst time used 200,000 times within the length of the outburst provides an error estimate of $\pm$0.46\,d. This is in aggreement with the recent period found in the \emph{Swift}-BAT light curve \citep{2010MNRAS.tmpL..63L}. 

\begin{figure}
	\includegraphics[width=0.95\linewidth,height=0.65\linewidth,trim = 8mm 3mm 0mm 6mm, clip]{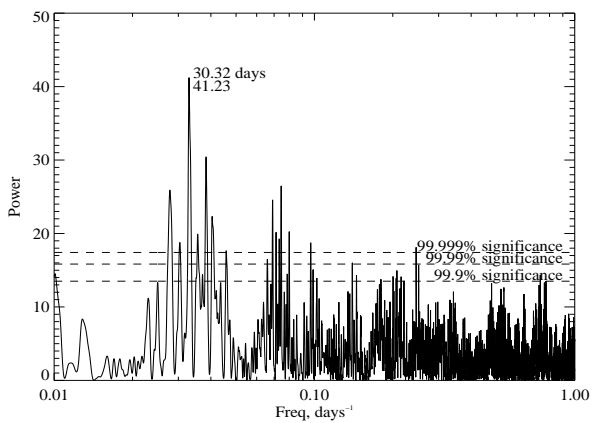}
	\caption{Periodogram for the IBIS light curve showing a period detection at 30.32 days.  The side lobes to the period can be shown to be due to the window function of the light curve, while the lower significance peak at $\sim$0.07 days$^{-1}$ is an alias of the main peak.}
	\label{figure:IBISPer}
\end{figure}

\section{Periodicity and Spectral Analysis}
\label{section:period}
\begin{figure}

	\includegraphics[width=0.97\linewidth,trim = 9mm 8mm 13mm 12mm, clip]{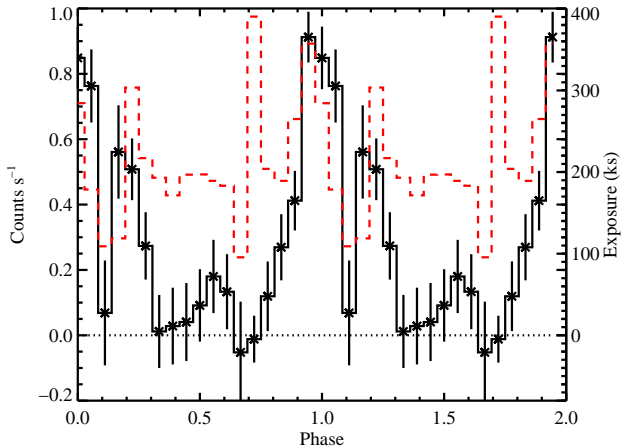}
	\caption{IBIS 18-60\,keV light curve folded on the discovered 30.32\,day period with the exposure for each bin overlaid (dashed).  An ephemeris of MJD 53736.2 has been determined to place the peak of the emission at phase 0.0.}
	\label{figure:explc}
\end{figure}
\begin{figure}
	\includegraphics[width=0.94\linewidth,trim = 9mm 8mm 0mm 6mm, clip]{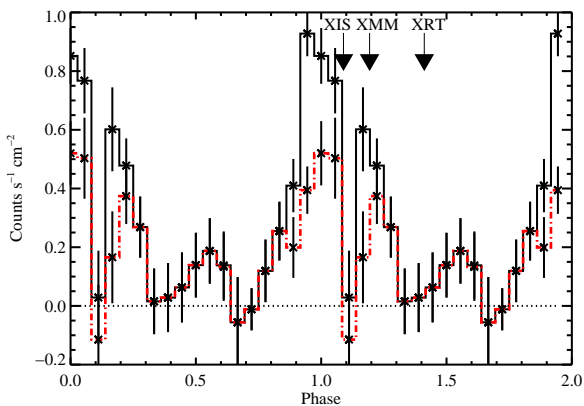}
	\caption{Folded IBIS light curve (solid line) with outbursts removed (dashed line).  Arrows indicate the phases at which follow up X-ray observations have been made by different X-ray instruments.}
	\label{figure:lcburst}
\end{figure}

\begin{figure}
	\includegraphics[width=0.95\linewidth,height=0.66\linewidth,trim = 8mm 3mm 0mm 6mm, clip]{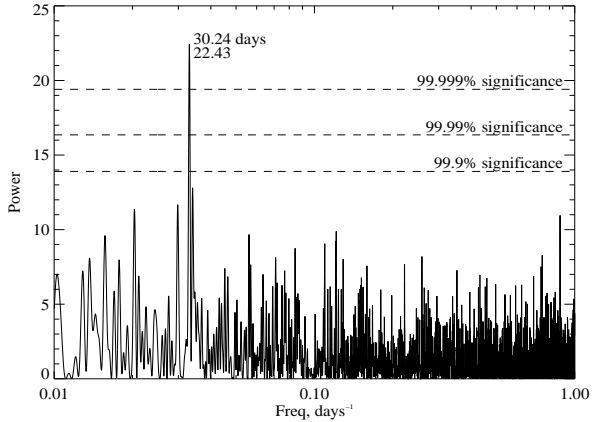}

	\caption{Periodogram for Swift orbit light curve showing a significant period at 30.23 days.\vspace{-12pt}}
	\label{figure:SwiftPer}
\end{figure}

The science window light curve (containing 2911  independent data points) was searched for signs of periodicity using the Lomb-Scargle periodogram method \citep{1976Ap&SS..39..447L, 1982ApJ...263..835S, 1989ApJ...343..874S}.  The Lomb-Scargle periodogram is shown in Figure~\ref{figure:IBISPer} with a clear peak evident at a frequency of 0.033 days$^{-1}$, corresponding to a period of 30.32 days, agreeing with the value determined by the PDM.  The significance of the peak was confirmed by performing a randomization test \citep{Hill:2005jy}; the flux points of the light curve are randomly re-ordered and a new Lomb-Scargle periodogram generated, the distribution of the power of the largest peak is an indicator of the significance of detection.  From 200,000 simulations we estimate the 99.9\%, 99.99\% and the 99.999\% significance levels which are shown in Figure~\ref{figure:IBISPer}; note that the maximum power achieved in any of the randomized light curves was $<$22.

The error on the identified period was estimated using a Monte-Carlo simulation. Each flux measurement was adjusted using Gaussian statistics within its individual error estimate to generate a simulated light curve of the source. The corresponding periodogram was produced and the location of the maximum peak near the frequency of 0.033 days$^{-1}$ was recorded. From 200,000 simulations we estimate the period and its error to be 30.32$\pm$0.02 (1$\sigma$) days.

Figure~\ref{figure:explc} shows the phase folded light curve using a period of 30.32 days and using an ephemeris of MJD 53736.2 to centre the fold at phase 0.0. Assuming any enhancement to the emission to coincide with the neutron stars closest approach to the donor, we consider phase 0.0 to be periastron. To see the effect of the outbursts on the phase folded curve a new fold was performed excluding all of the outbursts from Table~\ref{table:oldbursts}.  This second phase folded light curve is shown in Figure~\ref{figure:lcburst}, and shows no substantial difference in shape, demonstrating that there is an underlying periodicity in the flux aside from the outbursts when the entire light curve is considered.  The public light curves for IGR J16465--4507 from the \emph{Swift}-BAT mission were also analyzed using the Lomb-Scargle periodogram with the same tests used on the  \emph{INTEGRAL} light curve to see if there was any indication of the 30.32 day periodicity. The \emph{Swift} orbit light curve showed a period of 30.23$\pm$0.04\,d with a power of 22.43 (Fig.~\ref{figure:SwiftPer}), compatible with the period found in the \emph{INTEGRAL} light curve. 

There is some evidence for an eclipse (Fig.~\ref{figure:explc}) in the phase folded \emph{INTEGRAL} light curve.  However, this cannot be completely disentangled from low exposure in the relevant phase bins and the statistical significance of this dip is low.  The feature is also not apparent in the \emph{Swift} phase folded light curve leading to the conclusion that this is an exposure artifact and not a real feature.

The two outbursts with the highest significance in Table~\ref{table:oldbursts} (2 and 3) are shown in the inset of Figure~\ref{figure:fulllc}. Such IBIS data have been unsuccessfully searched for pulsations. The brightest flares cannot be seen by \emph{JEM-X} either on science window timescales  or in deep mosaics in the 4-20\,keV energy range, where the science windows exposure of 2.8\,ks was not sufficiently long to detect the source. This results in an upper limit for the flux during the outburst of 8$\times10^{-12}$\,erg\,cm$^{-2}\,$s$^{-1}$. Only one \emph{Swift}-XRT observation is available with only 69 events in 2\,ks and so it has not been possible to extract a meaningful spectrum and to search for pulsations. However this does give an estimated 0.2-10 keV absorbed (unabsorbed) flux of 9.1$\times10^{-12}$\,erg\,cm$^{-2}$\,s$^{-1}$ (4.9$\times10^{-11}$\,erg\,cm$^{-2}$\,s$^{-1}$), based on a power law index of 1.0, a Galactic n$_{H} =$ 2.12$\times$10$^{22}$ and an intrinsic n$_{H} =$ 6$\times$10$^{23}$. The flux has also been measured by \emph{XMM-Newton} as 3$\pm$1$\times10^{-12}$\,erg\,cm$^{-2}\,$s$^{-1}$ in the 4-10\,keV energy range \citep{2004ATel..336....1H} and as $8.95^{+0.2}_{-3.1}\times10^{-12}$\,erg\,cm$^{-2}\,$s$^{-1}$ in the 0.2-10\,keV energy range using {\it Suzaku}-XIS \citep{2009ApJ...699..892M}. These measurements approximately follow the drop in flux in the phase folded light curve (Fig.~\ref{figure:lcburst}).

Spectra for the two outbursts with the highest significance detection (n. 2 and 3 in table 1) have been extracted from the IBIS data. The 18--60 keV spectrum of the discovery outburst (n.3) is reasonably fit by a bremsstrahlung model 
($\chi^{2}_{\nu}$=0.6, 15 d.o.f., kT=18.6$^{+5}_{-3.5}$ keV, flux equal to 
1.2$\times10^{-10}$\,erg\,cm$^{-2}$\,s$^{-1}$) or alternatively by a power  law ($\chi^{2}_{\nu}$=0.7, 15 d.o.f, $\Gamma$ = 3.0$\pm$0.3). 
Similarly, the newly discovered outburst (n.2) is also reasonably fit by a bremsstrahlung ($\chi^{2}_{\nu}$=1.2, 15 d.o.f., kT=29$^{+11}_{-7}$ keV, 
flux equal to 9.5$\times10^{-11}$\,erg\,cm$^{-2}$\,s$^{-1}$) or by a power law ($\chi^{2}_{\nu}$=1.24, 15 d.o.f.,$\Gamma$ = 2.5$\pm$0.3). 
It is evident that both outbursts have a similar soft spectrum and a comparable average flux of activity, furthermore 
the reported  spectral shapes are very similar to those seen during outburst activity of other SFXTs 
(e.g. \citealt{2008A&A...487..619S, 2007A&A...467..249S}). 

Taking into account the 18--60\,keV bremsstrahlung spectral shape of both previous outbursts, 
we extrapolated their 0.2--10\,keV soft X-ray flux obtaining a value of $\sim$3$\times10^{-10}$\,erg\,cm$^{-2}$\,s$^{-1}$.  
To date the lowest and highest measured unabsorbed soft X-ray fluxes are equal to $\sim$3.6$\times10^{-12}$\,erg\,cm$^{-2}$\,s$^{-1}$ \citep{2004ATel..336....1H} 
and $\sim$4.9$\times10^{-11}$\,erg\,cm$^{-2}$\,s$^{-1}$ (this paper)  so the source displays a dynamical range from 6 to 83.
As for the hard X-ray band, if we compare the 22--50 keV quiescent flux ($\sim$ $7.7\times10^{-12}$\,erg\,cm$^{-2}$\,s$^{-1}$, \citet{Walter:2007ww}) 
to  the outbursts' fluxes as  listed in table 1, then the dynamical range is from 27 to 82. We note that such values 
are significantly higher than that recently  reported ($<$10) by \citet{2010MNRAS.tmpL..63L}
as based on 54 months BAT monitoring (15--50 keV) during which no evidence for flaring activity was found. 
However, we point out that such  apparent  dynamical range incongruences can be explained by the different 
BAT and IBIS instrumental capabilities. IBIS is particularly suited to detect short outbursts and so to reveal the dynamic 
range of fast X-ray transient sources, such as IGR J16465--4507,  thanks to its  higher instantaneous sensitivity 
for shorter observation lengths (i.e. science windows, $\sim$ 2000 s duration) which match very well the duration 
of the flares. BAT is more suited to long-term monitoring  thanks to its 
more continuous coverage but could eventually miss the detection of short flares which, if close to the IBIS sensitivity, 
are very likely  not detectable by BAT due to  its poorer instantaneous  sensitivity. It is worth noting that IBIS and BAT show to be complementary and 
their  combination is  important in investigating the  properties of SFXTs.

In order to test the recurrence of emission at periastron we carry out a recurrence analysis \citep{Bird:2009yq}.  For this test we calculate the light curve significance for a region of 4 days either side of the points considered as periastron and apastron (approximately 25\% of the orbit).  This will highlight if we see a low level emission at periastron passages compared to apastron. Comparing the distributions of significances for apastron and periastron using a Kolmogorov-Smirnov test we find a probability of 74\% that they have been drawn from different distributions. 
\vspace{-12pt}
\section{Discussion}

The orbital characteristics and stellar properties for IGR J16465--4507 are very similar to those of SAX J1818.6--1703 \citep{Bird:2009yq, 2009A&A...493L...1Z}.  The peak flux from the phase folded light curve with the outburst removed of SAX J1818.6--1703 as determined by \citet{Bird:2009yq} is $\sim$2.2\,c\,s$^{-1}$ in the ISGRI 18-60\,keV band.  Comparing this to the peak flux of  $\sim$0.55\,c\,s$^{-1}$ from the same light curve for IGR J16465--4507 shows that IGR J16465--4507 is $\sim$4 times weaker if we assume that these systems are very similar.  The difference in flux is compatible with the known distance measurements for both sources.  However, SAX J1818.6--1703 shows much less absorption  with $N_H = (6.0\pm0.7)\times10^{22}\,$cm$^{-2}$ compared to $N_H \sim6\times10^{23}\,$cm$^{-2}$ for IGR J16465--4507. 

Since masses and periods are known from the spectral type \citep{2000asqu.book..381D} and assuming that the compact object is a neutron star of mass 1.4$M_{\sun}$, Kepler's third law gives the semi-major axis for the system of 121.9-126.3$R_{\sun}$.  As we see no evidence for Roche-lobe overflow in the system the eccentricity of the system must be $<$0.8 for the smallest possible donor star size and $<$0.6 for the largest donor star size of 26$R_{\sun}$.

It is important to point out that whereas classical SFXTs  have a remarkable dynamical range ($>$100), 
we found for IGRJ16465--4507 significantly  lower values (in the range $\sim$30-80). Although in this regard IGR J16465--4507 cannot be considered a 
classical SFXT, still its dynamical range  deserves some attention since it is greater than that of  classical persistent variable 
supergiant HMXBs ($<$20). Our findings on the source temporal behaviour and its dinamical range are  in agreement 
with the results reported by Walter \& Zurita Heras (2007) and strongly support their idea that IGRJ16465--4507
is an intermediate SFXT system, much like few other similar cases reported in the literature \citep{Walter:2007ww}.

\begin{figure}
\includegraphics[width=0.95\linewidth,trim = 9mm 3mm 0mm 6mm, clip]{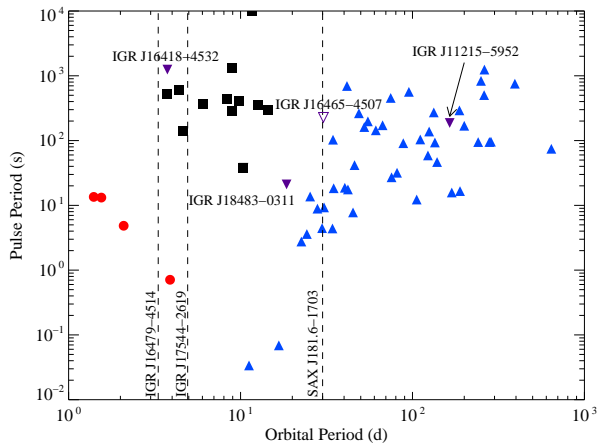}
	\caption{Corbet Diagram \citep{2006ATel..779....1C} showing the location of IGR J16465--4507 (clear inverted triangle) compared with other SFXTs (purple inverted triangles).  Be X-ray transient systems (triangles), Wind accretors (boxes) and Roche-lobe accretors (circles) are shown for reference.}
	\label{figure:corbet}
\end{figure}

As this system has both the orbital period and spin period known then it may be plotted on the Corbet diagram (Fig.~\ref{figure:corbet}).  This system resides between the regions dominated by wind accreting systems and Be X-ray transient systems, much like IGR J18483--0311 \citep{2007A&A...467..249S} and SAX J1818.6--1703.  When looked at in the context of the other SFXTs, IGR J16465--4507 sits in the middle of the class.  This implies a longer orbit than typical wind accreting supergiant systems, but without evidence for accretion from a circumstellar disk, as in Be X-ray binaries.  Our measurement of the orbital period in this system, combined with bursting behaviour of the source and constraints on the eccentricity of the orbits support the picture of accretion from a clumpy wind \citep{Negueruela:2006lt, Ducci:2009yq}. The underlying modulation in the phase folded light curve can be explained by the eccentricity of the system as with IGR J17544--2619 \citep{2009MNRAS.399L.113C}.  However, as shown in figure~\ref{figure:lcburst}, an underlying periodic emission, around periastron, does appear to be present in the phase folded light curve even when bright outbursts have been subtracted.

\section{Conclusions}

IGR J16465--4507 has been shown to exhibit an $\sim$30 day periodicity in both the {\em INTEGRAL}-IBIS and {\em Swift}-BAT light curves. This period is attributed to the orbital period of the system and corresponds to a semi-major axis of 121.9-126.3$R_{\sun}$.  This system is remarkably similar to SAX J1818.6--1703, except that IGR J16465--4507 is known to be further away, helping explain the lack of underlying emission that can only be seen in the phase-folded light curve of the entire {\em INTEGRAL} data set and not in the recurrence analysis. A low dynamic range is observed, which is higher than that of a typical supergiant HMXB. This would suggest that IGR J16465--4507 is an intermediate SFXT and that these may form a continuum ranging from typical supergiant HMXBs to the extreme SFXTs.

\section*{Acknowledgments}
Based on observations with {\em INTEGRAL}, an ESA project funded by member states (especially the PI countries: Denmark, France, Germany, Italy, Switzerland, Spain), Czech Republic and Poland, and with the participation of Russia and the USA.  A. B. Hill acknowledges support from the European Community via contract ERC-StG-200911. A. Bazzano and V. Sguera acknowledge support vai A.S.I. contract I/008/07.  \emph{Swift}-BAT transient monitor results provided by the  \emph{Swift}-BAT team.

\label{lastpage}

\end{document}